\documentclass{ws-procs9x6}
\newcommand{\Py}{{\sc Pythia}}
\begin{document}

\title{Bose Einstein Correlations in the Lund Model for Multijet Systems}

\author{Sandipan Mohanty}

\address{The Department of Theoretical Physics, \\
S\"olvegatan 14 A, \\ 
223 62 Lund, Sweden.\\ 
E-mail: sandipan@thep.lu.se}


\maketitle

\abstracts{The   interference   based   analysis  of   Bose   Einstein
Correlations in  the Lund  Model has hitherto  been limited  to simple
strings without gluonic excitations.  A new fragmentation method based
on  the Area  Law in  the Lund  Model allows  such an  analysis  to be
extended to multigluon strings.}

\section{Introduction}
The  Bose Einstein  effect  or  the enhancement  of  the two  particle
correlation  function for  identical bosons  with very  similar energy
momenta is well known in hadronic interactions. Since hadronisation is
mostly  described  through  phenomenological  models and  Monte  Carlo
simulations,  which  are based  on  classical probabilistic  concepts,
quantum  mechanical effects  such  as the  Bose Einstein  Correlations
(BEC) pose a problem.

In the event generator \Py, where hadronisation is handled through the
Lund  string   fragmentation  model,   this  effect  is   mimicked  by
introducing an attractive interaction  between identical bosons in the
final state.   The purpose behind  this is to parametrise  the effect,
rather than to provide a physical model for it.

A  physical model  for describing  the  BEC effect  within the  string
fragmentation scenario  was developed by Andersson and  Hofmann in [1]
which  was later  extended by  Andersson  and Ringn\'er  in [2].  They
showed that associating an amplitude with the decay of a string into a
set of hadrons  in the Lund Model leads  to interference effects which
enhance the probability for identical bosons forming a shade closer in
the  phase space than  what would  be expected  in a  purely classical
treatment, and identical fermions a shade farther appart.

But   their   formulation  was   limited   to   the  simplest   string
 configuration,  i.e.,  a string  stretched  between  a  quark and  an
 antiquark  with  no  gluonic  excitations.   Comparison  with  direct
 experimental data on BEC was not feasible, since a proper description
 of  the properties  of  hadronic jets  requires  parton showers,  and
 subsequent  fragmentation of  multigluon strings.   Even  though \Py~
 implements one approach  towards multigluon string fragmentation, the
 interference based  model for Bose  Einstein effect of  Andersson and
 Ringn\'er   could  not   be   extended  to   the  multigluon   string
 fragmentation scheme in \Py.

 Recently, an  alternative way to  fragment the multigluon  string has
been developed in  [3].  Unlike the approach in  \Py, this method does
not try to  follow the complicated surface of  a multigluon string. It
is based on the observation that  the string surface is a minimal area
surface in  space-time, and hence  it is completely determined  by its
boundary.  An attempt was  made to reformulate string fragmentation as
a process  along this boundary, called the  ``directrix''.  The result
was   a  new   scheme  for   string  fragmentation,   with   a  simple
generalisation  to multigluon strings.   This method  of hadronisation
has  been implemented  in an  independent Monte  Carlo  routine called
``ALFS''       (for       ``Area       Law      Fragmentation       of
Strings'')\footnote{available on request from the author.}.

Particle distributions from  ALFS are in agreement with  those of \Py~
on the  average, but  there are differences  at an exclusive  event to
event basis, which may show up in higher moments of the distributions.
It was also  understood that the interference based  model for the BEC
effect can be extended to multigluon string fragmentation in ALFS.

In  Sec.~\ref{alfs_frag}   this  new  fragmentation   scheme  will  be
summarised very briefly.  A brief  description of the basic physics of
the   interference   based   approach    to   the   BEC   appears   in
Sec.~\ref{lund_bec}.  In Sec.~\ref{cc} the concept of coherence chains
will be introduced  which allows the extension of  the analysis of BEC
in the  Lund Model to  multigluon strings.  Finally,  some preliminary
plots  obtained   by  using  this  method  to   analyze  two  particle
correlations will be presented in Sec.~\ref{results}.

\section{String Fragmentation as a process along the directrix}
\label{alfs_frag}
\begin{figure}[ht]
\centerline{\epsfxsize=1.9in\epsfbox{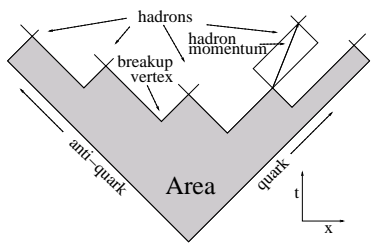}}   
\caption{Fragmentation of a String without gluonic excitations in the Lund Model. \label{area}}
\end{figure}

We recall that  the probability for the formation of  a set of hadrons
from a  given set of partons  in the Lund  Model, is given by  what is
known  as the ``Area  Law''. It  states that  this probability  is the
product of the final state phase space and the negative exponential of
the  area spanned  by the  string before  it decays  into  the hadrons
(cf. Figure \ref{area}):

\begin{eqnarray}
\label{arealaw}
dP_n(\{p_j\};P_{tot})=\prod_{j=1}^n   N_j  d^2p_j  \delta(p_j^2-m_j^2)
\delta( \sum_{j=1}^n p_j - P_{tot}) \exp(-b A)
\end{eqnarray}

An  iterative process based  on the  result in  Eq.~(\ref{arealaw}) is
fairly straight  forward to construct  for systems without  gluons. In
the Lund Model,  gluons are thought of as  internal excitations on the
string.   A  string  with  many such  excitations  traces  complicated
surfaces consisting of a large number of independent planar regions in
space--time.   One example  can  be seen  in Figure~\ref{ginterch}  in
Sec.~\ref{lund_bec}, which  illustrates the  surface of a  string with
just  one  gluon.   Calculating  the  energy momenta  of  the  hadrons
resulting  from  a  decay  of  strings  with  many  gluons  is  rather
difficult.

But since the world surface of  a string is a minimal area surface, it
has many  important symmetry properties  which may be  exploited while
considering  its decay  into a  set of  hadrons. Minimal  surfaces are
completely specified  by their  boundaries. For a  string in  the Lund
Model, this  boundary, called the ``directrix'', is  the trajectory of
the  quark  or the  antiquark  (one of  the  end  points).  Since  the
directrix determines  the string surface, it is  possible to formulate
string fragmentation  as a  process along the  directrix, as  shown in
[3].

\begin{figure}[ht]
\epsfxsize=10cm  
\epsfbox{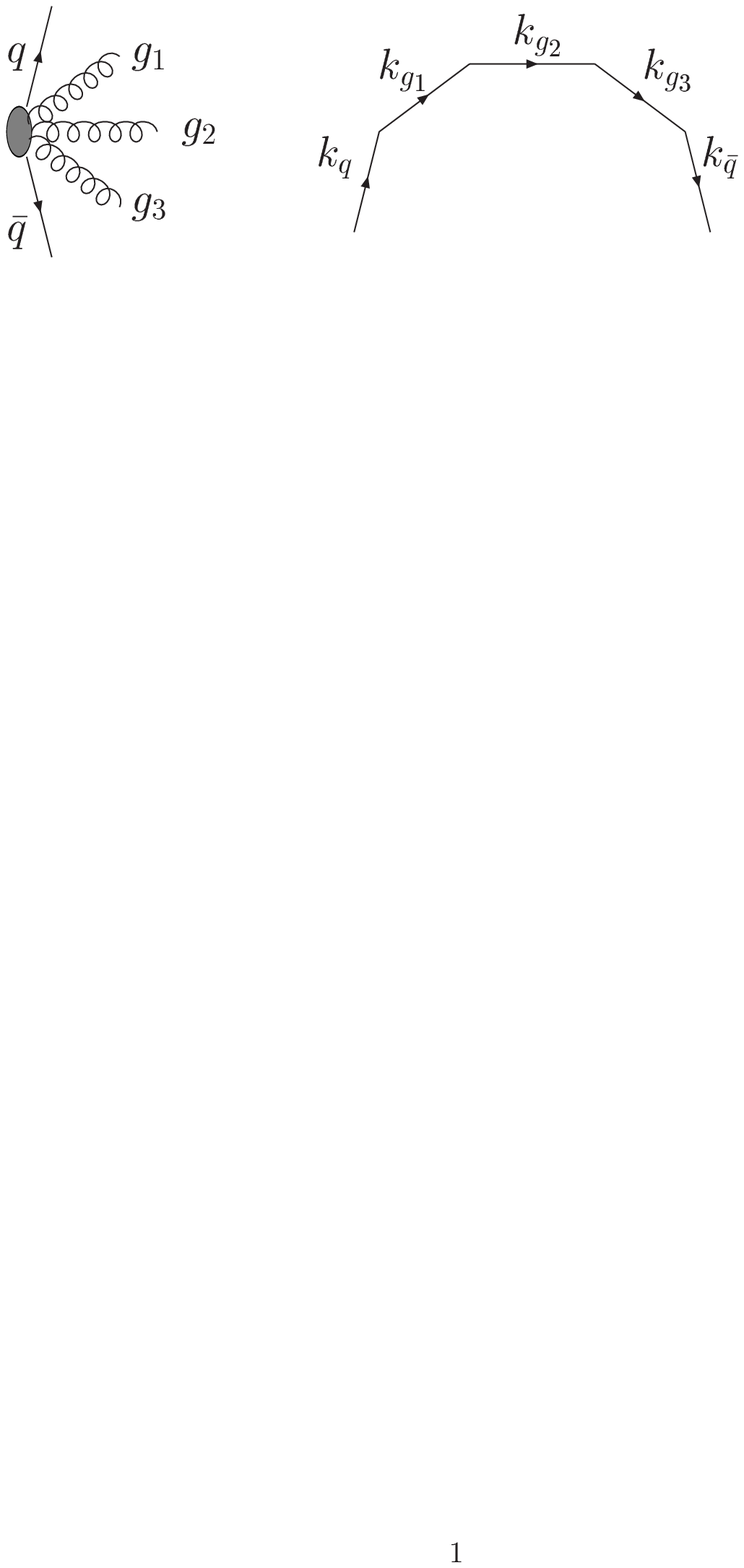}
\caption{Schematic representation of the directrix for a configuration
with a few resolved partons. \label{dir_constr}}
\end{figure}

The directrix  for a string,  which can be  thought to originate  at a
single  point  in space--time,  is  particularly  simple  and easy  to
visualize.   This   curve   can   be  constructed   by   placing   the
energy-momentum vectors of  the partons one after the  other in colour
order as shown (schematically) in Figure~\ref{dir_constr}.

\begin{figure}[ht]
\centerline{\epsfxsize=3.5in\epsfbox{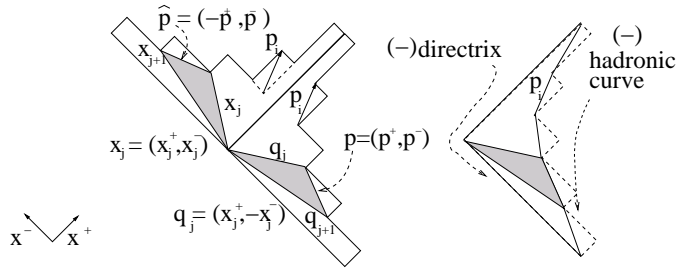}}
\caption{One  possible way  to  partition the  area  of a  fragmenting
string  into congributions for  each hadron.  It shows  the connection
between the  area in the area  law and the area  between the directrix
and the hadronic curve.\label{partition}}
\end{figure}

The fragmentation process developed in  [3] identifies the area in the
area  law with  the  area  between the  directrix  and the  ``hadronic
curve''\footnote{The string constant $\kappa$,  or the energy per unit
length in the string, will be  set to unity}, i.e., the curve obtained
by  placing the  hadron energy  momenta one  after the  other  in rank
order.  The  area  used  in  the  area law  can  be  partitioned  into
contributions  from the  formation of  each hadron  in  many different
ways.  Figure~\ref{partition} shows  one possible partitioning where a
triangular region  is associated with  one particle (shaded  region in
the upper left  part of the figure). This  figure also illustrates the
connection between the area  in Figure~\ref{area} and the area between
the  directrix and the  hadronic curve.   The upper  left part  of the
figure  shows  the  same  set  of  breakup  vertices  and  hadrons  as
Figure~\ref{area}.  The vectors $q_j$ in  the lower half of the figure
are obtained from the vertex vectors $x_j$ by inverting one light-cone
component  of $x_j$, and  are ``dual''  to the  vectors $x_j$  in this
sense.  They represent the  energy momentum  transfer between  the two
parts  of the  string  formed because  of  the breakup  at $x_j$.  The
triangular  regions   in  the  upper   part  of  the  figure   can  be
geometrically mapped to the triangular  regions in the lower part. But
the sum of the triangular areas  in the lower part is the area between
the directrix and  the hadronic curve whereas in the  upper part it is
the  area  as   used  in  Eq.~(\ref{arealaw})(ignoring  a  dynamically
uninteresting  constant  contribution  of  $\frac{1}{2}m^2$  for  each
hadron of mass $m$).

The   hadronisation  process   in  ALFS   associates   a  quadrangular
``plaquette'',  bounded  by the  hadron  energy  momentum vector,  two
'vertex' vectors,  and a  section of the  directrix, with  the hadron.
These  plaquettes  are  not  simple  geometrical  projections  of  the
triangular areas shown in  Figure~\ref{partition}, but their areas are
related in such a  way that the sum of the areas  of the plaquettes is
the  same as  the sum  of  the areas  of the  triangles. The  'vertex'
vectors in ALFS indeed do correspond to the space time locations where
quark antiquark pairs form  along the string during fragmentation, for
a flat string.   But in a more general context, it  is better to think
of them as somewhat more complex dynamical variables.

\begin{figure}[ht]
\centerline{\epsfxsize=1.6in\epsfbox{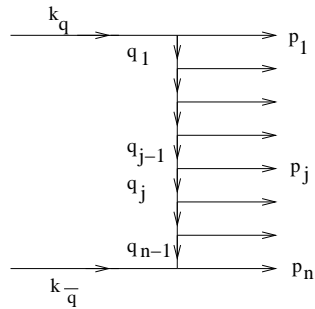}}
\caption{Lund Model  can be  thought of in  terms of a  ladder diagram
involving energy momentum exchanges. \label{ladder}}
\end{figure}

String  fragmentation (especially  as  implemented in  ALFS) could  be
thought of in terms of energy momentum transfer or ``ladder'' diagrams
like in  Figure~\ref{ladder}.  A quark momentum $k_q$  branches into a
hadron  momentum $p_1$ and  an energy  momentum transfer  $q_1$, which
then branches  into a  hadron vector $p_2$  and a new  energy momentum
transfer $q_2$,  and so  on. At each  stage the hadron  momentum forms
from the energy  momentum transfer vector comming into  that stage and
another independent vector which serves to define a longitudinal plane
in space--time. This other vector is just the anti--quark vector for a
flat string.  More generally it is a section of the directrix.
\vspace{-7mm}
This completes our brief overview of string fragmentation in ALFS. For
a detailed treatment and the  exact expressions the reader is referred
to [3].

\section{Physics of Bose Einstein Correlations in the Lund Model}
\label{lund_bec}

There is a formal similarity between the Area Law in Eq.~\ref{arealaw}
and  quantum  mechanical  transition  probabilities. And  even  though
hadronisation  is  a  quantum  mechanical process,  the  semiclassical
approach  in the  Lund Model  has been  very successful  in describing
experimental data. It is not impossible therefore, that the underlying
quantum mechanical process might  have an amplitude which when squared
resembles the area law.

In  [2] Andersson  and  Ringn\'er  argued that  one  can associate  an
amplitude of the form

\begin{eqnarray}
\label{amplitude}
\mu=e^{i(\kappa+ i b/2)A}
\end{eqnarray}

\begin{figure}[ht]
\centerline{\epsfxsize=1.9in\epsfbox{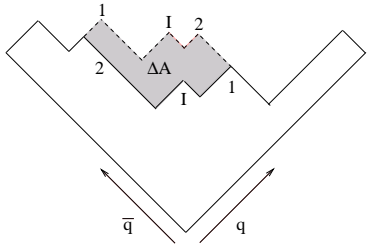}}   
\caption{Interchanging rank  order of  two identical particles  in the
final state  would require a different  set of breakup  vertices and a
different area under the string.  \label{interch}}
\end{figure}
where $\kappa$ is the string constant, with the decay of a string into
a set of hadrons.  This amplitude trivially reproduces the Area Law in
Eq.~\ref{arealaw}.   But it also  introduces interference  effects for
final states involving two or more identical particles, since for such
final states the string fragmentation model allows many different ways
to  produce  the  same final  state  from  a  given initial  state  as
illustrated  in  Fig~\ref{interch}.   The  figure shows  two  sets  of
breakup  vertices which  could lead  to the  same set  of  final state
particles in the  same flavour order. The particles  labeled ``1'' and
``2'', assumed  identical, have  interchanged rank orders  between the
two  schemes. The  two schemes  clearly involve  different  areas, and
hence     will    have     different    amplitudes     according    to
Eq.~(\ref{amplitude}).   This means  the total  squared  amplitude for
forming  such a  final state  (assuming there  are no  other identical
particles  in the  event) should  be  $|\mu|^2=|\mu_1+\mu_2|^2$, where
$\mu_1$ and  $\mu_2$ are  the amplitudes of  the two schemes  shown in
Figure~\ref{interch}. But a probabilistic Monte Carlo simulation would
assign a  probability $\mu_1^2+\mu_2^2$ with such a  state, which does
not account for  the interference term.  Thus, to  associate the right
probability with  the events  we may weight  this event with  an event
weight

\begin{eqnarray}
\label{evweight}
w=(1+2Re(\mu_1^{*}\mu_2)/(|\mu_1|^2+|\mu_2|^2))     
\end{eqnarray}
The result can be generalised to the case of many identical particles,
and  to include the  effect of  transverse momentum  generation during
hadronisation, as described in [2].

Treatment  of  string states  with  gluonic  excitations presents  new
problems. Since the  multiplicity of the events rises  with the number
of gluonic excitations, the  number of identical particles expected is
larger.   This  presents  a  computational problem.  More  importantly
though,  in this  case it  is  not always  possible to  find a  string
fragmentation  scheme  with  only  the  rank order  of  two  identical
particles   interchanged.   When  an   exchanged  scheme   exists  the
calculation   of  true  area   differences  and   transverse  momentum
contributions to  the amplitude is  rather involved, if  the exchanged
particles were originally produced in different planar regions.

\begin{figure}[ht]
\centerline{\epsfxsize=1.9in\epsfbox{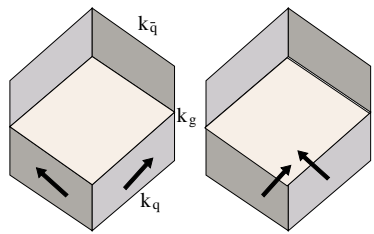}}
\caption{We  show here  the surface  traced by  a string  in  a system
consisting  of   a  quark,  a   single  gluonic  excitation,   and  an
antiquark.  Interchanging rank  order  of two  identical particles  in
different  string  planes seems  unnatural.  The interchanged  schemes
would  have  very  low  probabilities  to be  produced  during  string
fragmentation. It  may help to  think of the two  surfaces represented
here   like  two   chairs  facing   the  reader,   for  visualisation.
\label{ginterch}}
\end{figure}

But  in   string  fragmentation,  the  particle   energy  momenta  are
constructed from local momentum flow along the string world surface in
the  neighbourhood of  the breakup  vertices. Therefore,  most  of the
energy momentum of a hadron is along the local longitudinal directions
relative  to the  string. Figure~\ref{ginterch}  once again  shows two
identical particles  formed in different  regions in the  string.  But
this  time  they do  not  belong  to the  same  planar  region on  the
string. It is clear that the ``exchanged'' scheme (shown to the right)
would  be highly unlikely  to emerge  from this  string as  the energy
momenta  are no  longer  nearly aligned  with  the local  longitudinal
directions.

It was  mentioned earlier that  the fragmentation scheme in  ALFS does
not depend on  explicit representations of the string  surface such as
the one  in Figure~\ref{ginterch}. In  that approach, it  is sometimes
possible to  find another partonic  configuration which may  result in
the exchanged scheme as one  possible event. But if the partonic state
is held  fixed, such an exchange  would be improbable  for the reasons
just mentioned.

As a first approximation therefore,  it is reasonable to calculate BEC
on  multigluon strings  by  considering particle  permutations in  the
planar  regions of  the string  surface  and ignoring  the effects  of
exchange of particles  across gluon corners. But the  number of gluons
and the  size of planar  regions on the  string depend on  the cut-off
scale in the  parton cascade used to generate the  partons in an event
generator.   It  would  therefore  seem  that by  making  the  cut-off
sufficiently small we can make  the planar regions so small that there
would not be any instances of identical particles in one planar region
anywhere in  the event. To address  this, we introduce  the concept of
coherence chains.

\section{Coherence Chains}
\label{cc}
When  the cut-off  scale in  the ordering  variable  (gluon transverse
momentum, for  example) is  made small, softer  and softer  gluons are
resolved.   For  a   relatively  soft   gluon,  the   two   planes  in
Figure~\ref{ginterch} will  be only slightly inclined  with respect to
each other, and the exchanged scheme would not appear so unnatural. If
such exchanges are permitted in  ALFS, the new partonic states created
will not be outrageously different from the one we started with.

However,  parton  showers  are  probabilistic  in  the  Monte  Carlos.
Information   about   phases    involved   with   different   partonic
configurations  are ``lost''.   To analyse  permutations  of identical
hadrons across gluon corners, we need to consider interference effects
between results of hadronisation  from two slightly different partonic
configurations.  This  appears to be  problematic as we need  both the
phase  information  from  the   string  fragmentation  and  the  phase
information from the partonic stage while calculating the interference
terms and event weights.

Infrared  stability  of  string  fragmentation,  on  the  other  hand,
suggests that  the detailed properties  of the hadronic  states should
not  be extremely  sensitive to  gluon emission  around  hadronic mass
scales. In  a sense the  string state is  resolved at the  hadron mass
scales by  the fragmentation  process. One interesting  consequence of
this  was  observed  for  the  set  of hadrons  emerging  out  of  the
fragmentation of multigluon strings in ALFS.

The energy momenta  of the hadrons could be  collected into sets, such
that inside  each set, the  energy momenta are  aligned in a  plane in
space   time    upto   a   small   scale    in   transverse   momentum
fluctuations. This suggests that at least some aspects of the hadronic
phenomena might be insensitive to  the softest gluons generated by the
parton showers.

With an analysis  of BEC in mind we call these  groups of particles in
the final state as ``Coherence  Chains''. They describe the regions on
the string  over which  coherent interference effects  between hadrons
should be  considered. As  we have seen,  it seems quite  unnatural to
consider    symmetrisation   across    hard    gluon   corners,    cf.
Fig~\ref{ginterch},  whereas  symmetrisation  across  soft  gluons  is
necessary. The transverse momentum resolution scale used to define the
coherence chains  should be chosen such that  it distinguishes between
these situations.

The  approximation being  made  in  the analysis  of  BEC through  the
coherence chains  could be stated  as follows: we ignore  the possible
effects  on BEC,  of the  slightly different  amplitudes  of different
partonic  states which  may give  rise  to one  coherence chain  after
hadronisation. To calculate BE weights, we treat the hadronic state as
if it came from a simpler  string state which has only those planes in
it which are present in  the coherence chains.  Symmetrisation is then
carried  out separately  for  each plane  and  the squared  amplitudes
multiplied and a suitable event weight calculated.

The hadron energy  momenta are not directly altered  as in PYBOEI (the
BE subroutine in \Py), but different events receive different weights.

There  is a  tendency for  events  with higher  multiplicity to  yield
higher weights. Since multiplicity  is a function of gluonic activity,
it is not possible to retune parameters pertaining to hadronisation to
compensate  for  the multiplicity  dependence  of  weights, unless  we
associate   a  total  of   one  hadronic   state  for   each  partonic
configuration.   This  leaves  only  the possibility  of  a  rejection
weighting  on the hadronized  states based  on their  BE weights  in a
Monte  Carlo. This  procedure is  much  slower than  PYBOEI.  But  the
purpose  of this exercise  is to  provide a  physical picture  for the
phenomenon inside the Lund Model.
  
\section{Preliminary Results and Concluding Remarks}
\label{results}
The  interference based analysis  of BEC  in the  Lund Model  has been
extended so as to be  applicable to multigluon string fragmentation as
implemented  in ALFS.   Modules for  BEC calculations  have  also been
introduced  into  ALFS.  A  preliminary  analysis  shows the  expected
enhancement of the two particle correlation function at small momentum
differences.   For events  with a  few  prominent jets,  BEC tends  to
decrease with  the number of  jets if the total  $\lambda$-measure for
the strings is kept fixed, cf Figure 7.  No significant correlation is
seen between oppositely  charged pions, cf Figure 8.  A detailed study
of the properties of coherence chains, how they affect the analysis of
BEC and further studies of this model for BEC itself will be presented
elsewhere.

\begin{minipage}[t]{4.5cm}
\leftskip=-7mm
\rotatebox{270}{\centerline{\epsfxsize=4cm\epsfbox{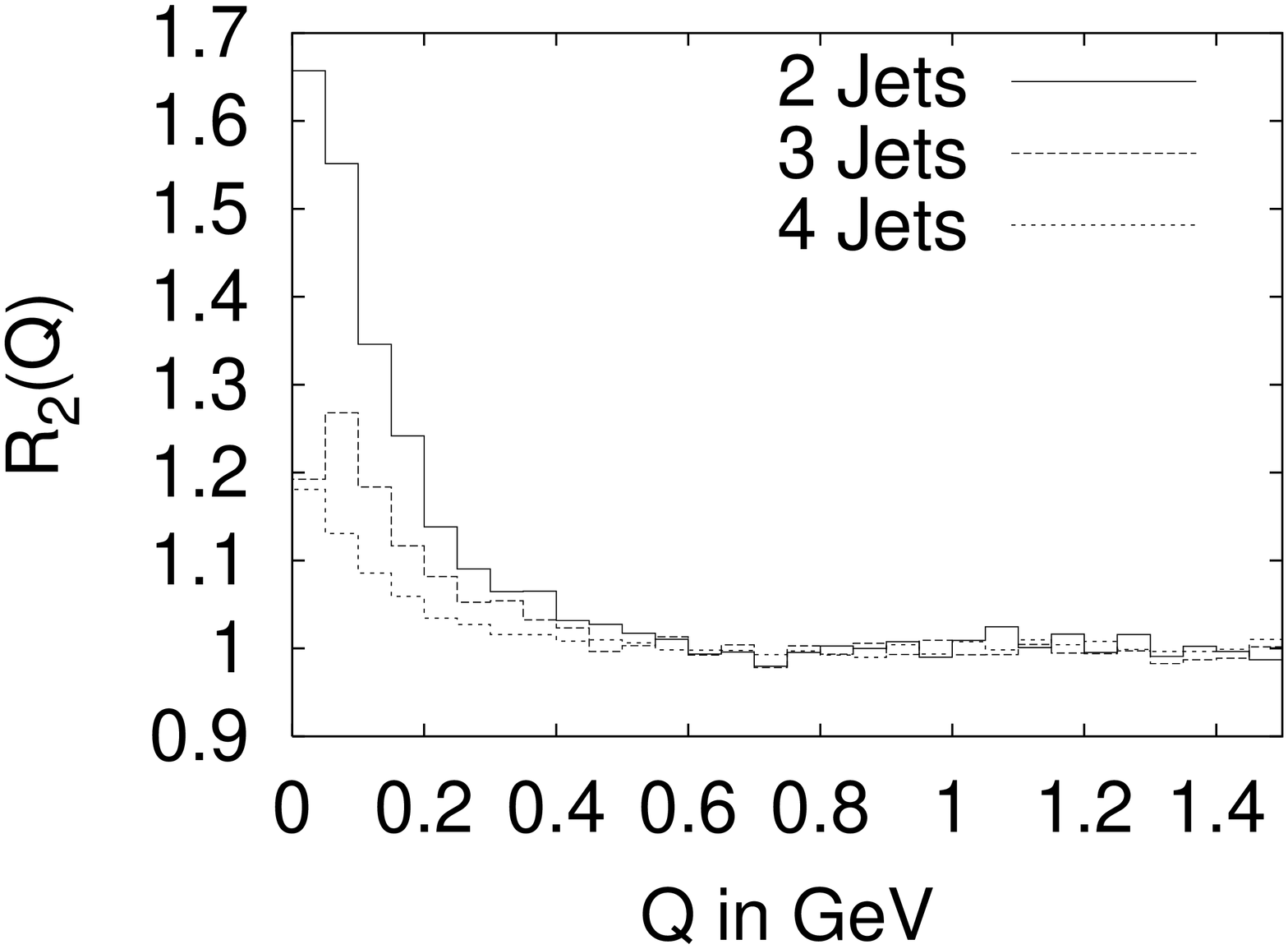}}}
\par
\end{minipage}
\hskip6mm
\begin{minipage}[t]{4.5cm}
\rotatebox{270}{\centerline{\epsfxsize=4cm\epsfbox{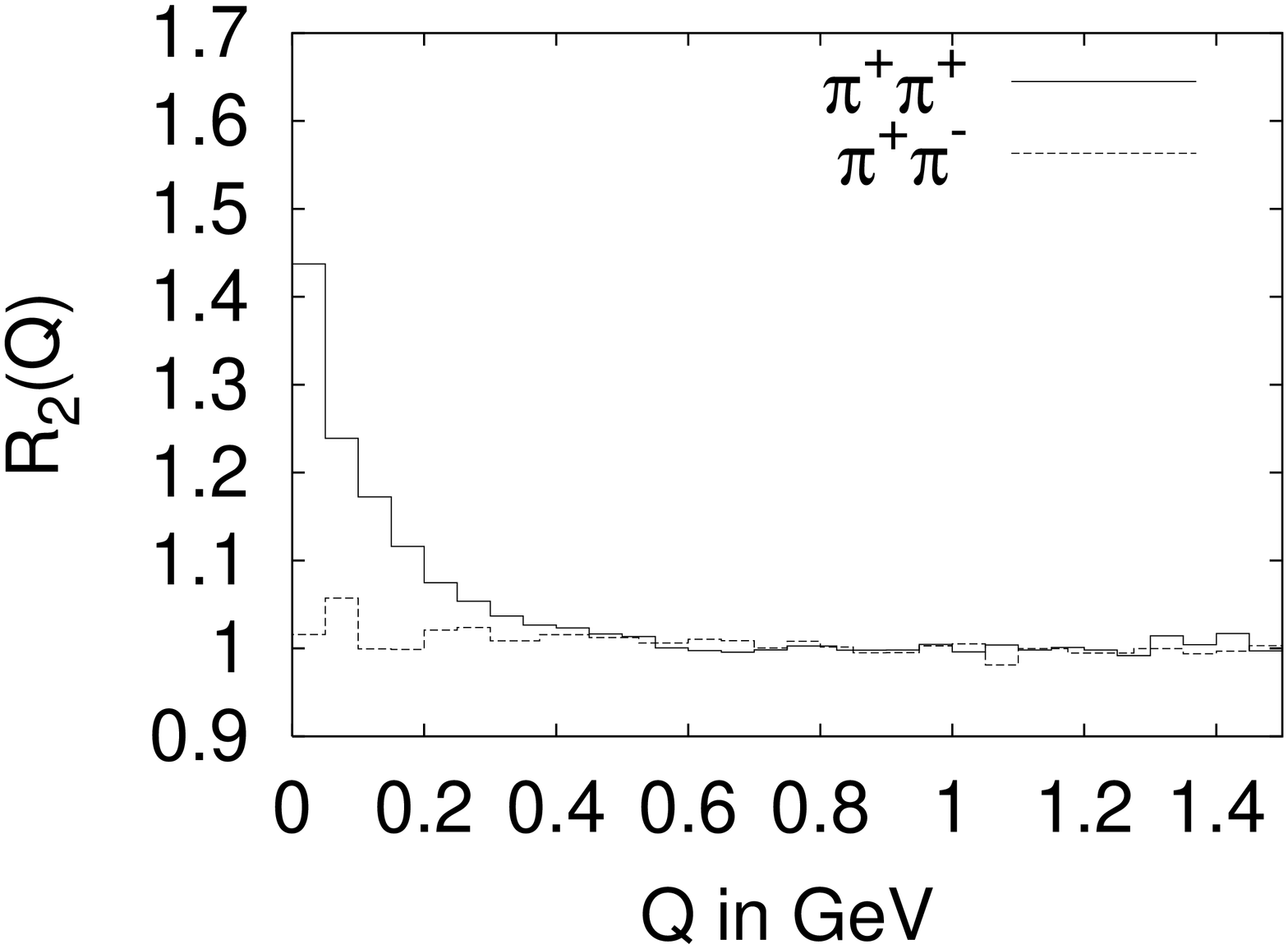}}}
\end{minipage}
\vskip 2mm
\begin{minipage}[t]{4.5cm}
{\small\baselineskip=10pt\noindent
\leftskip=-5mm
Figure 7. Two particle correlation function from ALFS for systems with few jets. ``String length'' or $\lambda$-measure was kept fixed. \par}
\end{minipage}
\hskip7mm
\begin{minipage}[t]{4.5cm}
{\small\baselineskip=10pt\noindent
Figure 8. This plot shows that no significant correlation effects are expected between oppositely charged pions in this model.\par}
\end{minipage}
\par
\vspace{7mm}

\section*{Acknowledgements}
This  project began  as  a  collaboration between  the  Late Prof.  Bo
Andersson, my collegue Fredrik  S\"oderberg and myself. Even though it
is still an unfinished project and new developments are being made, we
are indebted to Prof. Andersson for the numerous insights he provided,
while he was with us.

\end{document}